\documentclass[a4paper]{amsart} 
\usepackage{amsmath}
\usepackage{epsfig}
\usepackage{amssymb} 

\renewcommand{\i}{{\mathrm{i}}}

\begin{document}

\title{ Hyperelliptic Theta-Functions and Spectral Methods II} 
\author{J.~Frauendiener}
\address{Institut f\"ur Astronomie und Astrophysik, Universit\"at T\"ubingen,
Auf der Morgenstelle 10, 72076 T\"ubingen, Germany}
\email{joerg.frauendiener@uni-tuebingen.de}

\author{C.~Klein}
\address{Max-Planck-Institut f\"ur Physik, F\"ohringer Ring 6,
    80805 M\"unchen, Germany}
\curraddr{Max-Planck-Institut for Mathematics in the Sciences, 
Inselstr. 22-26, 04103 Leipzig, Germany}
    \email{klein@mis.mpg.de}
\date{\today}    

\begin{abstract}
   This is the second in a series of papers on the numerical treatment
   of hyperelliptic theta-functions with spectral methods.  A code for
   the numerical evaluation of solutions to the Ernst equation on
   hyperelliptic surfaces of genus 2 is extended to arbitrary genus
   and general position of the branch points.  The use of spectral
   approximations allows for an efficient calculation of all
   characteristic quantities of the Riemann surface with high
   precision even in almost degenerate situations as in the solitonic
   limit where the branch points coincide pairwise. As an example we
   consider hyperelliptic solutions to the Kadomtsev-Petviashvili and
   the Korteweg-de Vries equation. Tests of the numerics using
   identities for periods on the Riemann surface and the differential
   equations are performed. It is shown that an accuracy of the order
   of machine precision can be achieved.
\end{abstract}

\maketitle

\section{Introduction}

Solutions to integrable differential equations in terms of
theta-functions were introduced at the beginning of the seventies, see
\cite{algebro} for an account of the history. In general, they
describe periodic or quasi-periodic solutions. In contrast to the
well-known solitonic solutions in terms of elementary functions, the
theta-functions are special transcendental functions defined on some
Riemann surface. All quantities entering the solution are given in
explicit form via integrals which depend implicitly on the branch
points of the Riemann surface. A full understanding of the functional
dependence on these parameters seems to be only possible
numerically. Algorithms have been developed to establish such
relations for rather general Riemann surfaces as in~\cite{tretkoff84}
or via Schottky uniformization (see \cite{algebro,bobbor}), which have
been incorporated successively in numerical and symbolic codes,
see~\cite{deconinck01,deconinck03,dubrovin97,gianni98,hoeij94,seppala94} and
references therein (the first two references are distributed along with
Maple~6, respectively Maple~8, and as a Java
implementation~\cite{riemann}).

These codes are convenient to study theta-functional solutions on a
given surface. To establish the dependence of the characteristic
quantities of a Riemann surface on the branch points, an efficient
calculation of the periods is necessary. This is especially true if
the modular dependence of the theta-functions is of interest as in the
case of theta-functional solutions to the Ernst equation~\cite{ernst}
in \cite{prl2,Koro88}.

In \cite{numerik1}, henceforth referred to as I, we have presented a
code for the numerical treatment of hyperelliptic
solutions to the Ernst equation on
a surface of genus~2. The integrals entering the solution are
calculated by expanding the integrands as a series of Chebyshev
polynomials using a Fast Cosine Transformation in MATLAB and then
integrating the polynomials in an appropriate way. In the present
article, this code is extended to hyperelliptic surfaces of arbitrary
genus and of branch points in general position. As an example we
consider hyperelliptic solutions to the Kadomtsev-Petviashvili (KP)
equation and the Korteweg-de Vries (KdV) equation. The precision of
the numerical evaluation is tested by checking identities for periods
on Riemann surfaces and the differential equations. We show that an
accuracy of the order of machine precision ($\sim 10^{-14}$) can be
achieved for branch points in general position with 32 polynomials and
in the case of almost degenerate surfaces which occurs e.g.\ in the
well-known solitonic limit with at most 256 polynomials. Consequently
the solitonic limit can be carried out numerically with machine
precision which will be shown at the example of the 2-soliton solution
to the KdV equation. This makes it possible to study the parameter
space for hyperelliptic surfaces numerically.

This paper is a sequel to I which was devoted to genus 2 solutions to
the Ernst equation. We will only briefly repeat the methods already
outlined there and refer the reader to I for details. The article is
organized as follows: in section~\ref{sec:kp} we collect useful facts
on the KP and the KdV equation and hyperelliptic Riemann surfaces, in
section~\ref{sec:spectral} we summarize basic features of spectral
methods and explain our implementation of various quantities. The
calculation of the periods of the hyperelliptic surface is performed
together with tests of the precision of the numerics. The theta-series
is approximated as in \cite{deconinck03} as a finite sum. It is
checked numerically how well the theta-functional solution solves the
differential equation. In section~\ref{sec:examples} we present
several examples of hyperelliptic solutions to the KP and the KdV
equation in general cases and in almost solitonic situations.  In
section~\ref{sec:concl} we add some concluding remarks.

\section{KP and KdV equation and hyperelliptic Riemann surfaces}
\label{sec:kp}

The KP equation for the real valued potential $u$ depending on the
three real coordinates $(x,y,t)$ can be written in the form
\begin{equation}
    3u_{yy} + \partial_{x}(6uu_{x}+u_{xxx}-4u_{t})=0
    \label{eq:kp}.
\end{equation}
The completely integrable equation has a physical interpretation as
describing the propagation of weakly 
two-dimensional waves of small amplitude in shallow water as well as 
similar physical processes, see \cite{algebro}. We note that there is 
a variant of the KP equation having a different sign of the 
$u_{xxx}$-term which will not be considered here. 

Algebro-geometric solutions to the KP equation can be given on an 
arbitrary Riemann surface. Here we will only consider hyperelliptic 
surfaces $\Sigma_{g}$ of genus $g$ 
corresponding to the plane algebraic curve
\begin{equation}
    \mu^{2}=\prod_{i=1}^{g+1}(K-E_{i})(K-F_{i})=:K^{2g+2}-S_{1}
K^{2g+1}+S_{2}K^{2g}+\ldots
    \label{hyper1},
\end{equation}
where we have defined the symmetric (in the branch points) functions 
$S_{n}$, $n = 1, \ldots, 2g+2$.
We will concentrate on real surfaces with real branch points $E_{i}$, 
$F_{i}$. The branch points are ordered, $E_{1}<F_{1}<\ldots<E_{g+1}< 
F_{g+1}$. 

\begin{figure}[htb]
    \centering \epsfig{file=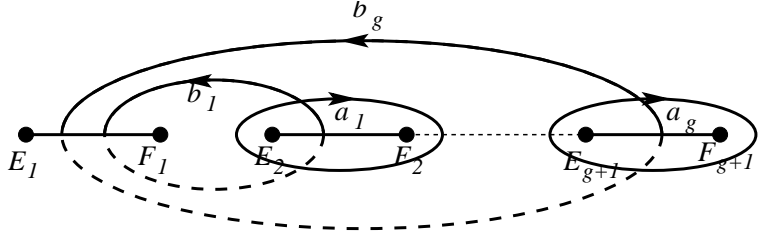,width=10cm}
    \caption{Canonical cycles.}
    \label{fig:cut-system}
\end{figure}

Hyperelliptic Riemann surfaces are important since they show up in the
context of algebro-geometric solutions of various integrable equations
such as KdV, Sine-Gordon and Ernst. Whereas it is a non-trivial problem to
find a basis for the holomorphic differentials on general surfaces
(see e.g.~\cite{deconinck01}), it is given in the hyperelliptic case
(see e.g.~\cite{algebro}) by
\begin{equation}
\left(   \frac{dK}{\mu}, \frac{KdK}{\mu},\ldots,
  \frac{K^{g-1}dK}{\mu} \right)
    \label{basis},
\end{equation}
which is the main simplification in the use of these surfaces.  We
introduce on $\Sigma_{g}$ a canonical basis of cycles $(a_{k},b_{k})$,
$k=1,\ldots,n$ as in Fig.~\ref{fig:cut-system}. The holomorphic
differentials $d\omega_k$ are another basis in the space of
holomorphic differentials which is normalized by the condition on the
$a$-periods
\begin{equation}
    \int_{a_{l}}^{}d\omega_{k}=2\pi \i \delta_{lk}.
    \label{normholo}
\end{equation}
This fixes the $d\omega_k$ uniquely given the system of cycles.  The
corresponding matrix of $b$-periods is given by $\mathbf{B}_{lk} =
\int_{b_{l}}^{}d\omega_{k}$. The matrix $\mathbf{B}$ is a so-called
Riemann matrix, i.e.\ it is symmetric and has a negative definite real
part.

The surfaces we consider here are real, i.e.\ they have an
anti-holomorphic involution $\tau: \Sigma_{g}\to \Sigma_{g}$ with
$\tau^{2}=1$. The basis of the homology is chosen to have the property
\begin{equation}
    \tau b_{j} = b_{j}, \quad \tau a_{j} = -a_{j}, \quad j = 1, 
\ldots, g
    \label{eq:real1}.
\end{equation}
This implies for the normalized holomorphic differentials $d\omega_{j}$ 
the reality condition 
\begin{equation}
    \tau d\omega_{j} = \overline{d\omega_{j}}
    \label{eq:real2}.
\end{equation}
The matrix of $b$-periods is real in this case.

The theta-function with characteristics corresponding to the curve
$\Sigma_{g}$ is given by
\begin{equation}
    \Theta_{\mathbf{p}\mathbf{q}}(\mathbf{x}|\mathbf{B})=
    \sum_{\mathbf{n}\in\mathbb{Z}^{g}}^{}\exp\left\{\frac{1}{2}
    \langle\mathbf{B}(\mathbf{p}+\mathbf{n}),(\mathbf{p}+\mathbf{n})
    \rangle+\langle\mathbf{p}+\mathbf{n},2\pi\i\,\mathbf{q}+\mathbf{x}
    \rangle\right\}
    \label{theta},
\end{equation}
where $\mathbf{x}\in \mathbb{C}^{g}$ is the argument and
$\mathbf{p},\mathbf{q}\in \mathbb{C}^{g}$ are the characteristics. We
will only consider half-integer characteristics in the following. The
theta-function with characteristics is, up to an exponential factor,
equivalent to the theta-function with zero characteristic (this is the
Riemann theta-function, denoted with $\Theta$) and shifted argument,
\begin{equation}
    \Theta_{\mathbf{p}\mathbf{q}}(\mathbf{x}|\mathbf{B})=
    \Theta(\mathbf{x}+\mathbf{B}\mathbf{p}+2\pi\i\, \mathbf{q})\exp\left\{
    \frac{1}{2}\langle\mathbf{B}\mathbf{p},\mathbf{p}
    \rangle+\langle\mathbf{p},2\pi\i\,\mathbf{q}+\mathbf{x} \rangle\right\}.
    \label{theta2}
\end{equation}
It has the periodicity properties 
\begin{equation}
    \Theta_{\mathbf{p}\mathbf{q}}(\mathbf{z}+2\pi \i\,\mathbf{e}_{j}) = 
    e^{2\pi ip_{j}}
    \Theta_{\mathbf{p}\mathbf{q}}(\mathbf{z}),
    \quad 
    \Theta_{\mathbf{p}\mathbf{q}}(\mathbf{z}+\mathbf{B}
    \mathbf{e}_{j})=
    e^{-2\pi \i\, q_{j}-z_{j}-\frac{1}{2}B_{jj}}
    \Theta_{\mathbf{p}\mathbf{q}}(\mathbf{z})
    \label{eq:periodicity},
\end{equation}
where $\mathbf{e}_{j}$ is the $g$-dimensional vector consisting of 
zeros except for a 1 in jth position.

The differentials of the second kind needed here
have a pole of order $l=1,2,3$ at
infinity. They will be denoted by $d\Omega_{l}$. The Laurent expansion
of the corresponding integrals is of the form
\begin{equation}
    \int_{\infty}^{P}d\Omega_{l}= k^{l}+o(1)
    \label{eq:second},
\end{equation}
where $k$ is the local parameter in the vicinity of infinity. The
singularity structure characterizes the differentials only up to an
arbitrary linear combination of holomorphic differentials. This
arbitrariness is eliminated by imposing the condition that all
$a$-periods vanish. The normalized differentials satisfy the reality
conditions $\tau d\Omega_{l}=\overline{d\Omega_{l}}$. The vectors of
$b$-periods of these differentials are denoted by
\begin{equation}
    U_{n}= \int_{b_{n}}^{}d\Omega_{1}, \quad V_{n}= 
\int_{b_{n}}^{}d\Omega_{2}, \quad W_{n}=\int_{b_{n}}^{}d\Omega_{3}
    \label{eq:speriods}.
\end{equation}
The periods are real under the above conditions. For later use we 
define the constant $c$ via the expansion of the integral 
$\Omega_{1}$, 
\begin{equation}
    \int_{\infty}^{P}d\Omega_{1}=k-\frac{c}{k}+O(k^{-2})
    \label{eq:c}.
\end{equation}

The differentials of the second kind $d\Omega_{i}$ are given
explicitly on $\Sigma_{g}$ by
\begin{equation}
  d\Omega_{1}=\frac{K^{g+1}-\frac{1}{2}S_{1}K^{g}+c_{1}K^{g-1}+\ldots+
c_{g}}{2\mu(K)}\,dK
    \label{eq:diff1},
\end{equation}
where the $c_{l}$ are determined by the condition of vanishing 
$a$-periods, 
\begin{equation}
    d\Omega_{2}=\frac{K^{g+2}-\frac{1}{2}S_{1}K^{g+1}-(\frac{1}{8}S_{1}^{2}-
\frac{1}{2}S_{2})
K^{g}}{\mu(K)}\,dK + \ldots
    \label{eq:diff2},
\end{equation}
and
\begin{equation}
    d\Omega_{3}=\frac32 \frac{K^{g+3}-\frac{1}{2}S_{1}K^{g+2}-(\frac{1}{8}S_{1}^{2}
-\frac{1}{2}S_{2})
  K^{g+1}-(\frac{1}{16}S_{1}^{3}-\frac{1}{4}S_{1}S_{2}+
\frac{1}{2}S_{3})K^{g}}{\mu(K)} dK +\ldots
    \label{eq:diff3},
\end{equation}
where $\ldots$ denotes the holomorphic differentials leading to vanishing 
$a$-periods.
For the constant $c$, we get 
\begin{equation}
    c = \frac{S_{1}^{2}}{8}-\frac{S_{2}}{2} +c_{1}
    \label{eq:cex}.
\end{equation}
It is well known that the $b$-periods of normalized differentials of
the second kind can be expressed in terms of expansions of the
holomorphic differentials in the vicinity of the singularity (see
e.g.~\cite{algebro}).  We write the holomorphic differentials in the
vicinity of some point $a$ in the form
\begin{equation}
    d\omega = \sum_{n=0}^{\infty}\mathbf{v}_{n}\frac{t^{n}}{n!}dt
    \label{eq:holexp},
\end{equation}
where $t$ is the local parameter in the vicinity of $a$. If we choose 
$a=\infty^{+}$ (the infinite point in the upper sheet), we have
\begin{equation}
    \mathbf{v}_{0}=\mathbf{U},\quad  \mathbf{v}_{1}=\mathbf{V}, \quad 
\mathbf{v}_{2}=\mathbf{W}/2
    \label{eq:bperiods}.
\end{equation}
Thus, the periods $\mathbf{U}$, $\mathbf{V}$ and $\mathbf{W}$ can be 
calculated as the $b$-periods of the differentials $d\Omega_{i}$, 
$i=1,2,3$, or via an expansion of the holomorphic differentials.

Solutions to the KP equation on the above Riemann surfaces are given 
by the generalization of the Its-Matveev formula for the KdV equation
(see e.g.\ \cite{mumford})
\begin{equation}
    u = 2\partial_{x}^{2}\ln 
\Theta(\mathbf{U}x+\mathbf{V}y+\mathbf{W}t+\mathbf{D})+2c
    \label{eq:solution},
\end{equation}
where $\mathbf{D}\in \mathbb{C}^{g}$ is an arbitrary real
vector. Because of (\ref{theta2}) and the second logarithmic
derivative in (\ref{eq:solution}), the vector $\mathbf{D}$ can always
be absorbed in the form of a real characteristic. By a coordinate
change $x\to x + tc/3$ one can change a solution $u$ to the KP
equation by $-2c$.

The reality of the argument of the theta-function implies that the
theta-function which is an entire function does not vanish. This can
be seen from the following argument for the elliptic case:
\begin{eqnarray}
    \Theta(x) & = & \sum_{m\in\mathbb{Z}}^{}\exp 
\left(\frac{1}{2}m^{2}B+mx\right)
    \nonumber  \\
     & = & 1+2\sum_{m\in\mathbb{N}}^{} \exp 
\left(\frac{1}{2}m^{2}B\right)\cosh(mx)>0
    \label{eq:regular}  .
\end{eqnarray}
A generalization of the argument to higher genus is straightforward. 

Probably the most elegant way to show that (\ref{eq:solution})
provides a solution to the KP equation is the use of Fay's trisecant
identity \cite{fay}. This is an identity for theta-functions which
holds for four arbitrary points on a Riemann surface, see
\cite{mumford} for details. Here we need the identity in the limit
that all four points coincide which leads to an identity for
derivatives of theta-functions. We introduce the directional
derivatives on $\mathbb{C}^g$ ($\nabla$ acts on $\mathbf{z}\in \mathbb{C}^g$)
\begin{equation}
    D_{a} = \mathbf{v}_{0} \cdot \nabla,\quad D_{a}'=\mathbf{v}_{1}
    \cdot \nabla, \quad  
D_{a}'' = \mathbf{v}_{2} \cdot \nabla
    \label{eq:dirder}.
\end{equation}
Then for an arbitrary  vector $\mathbf{z}$ the following identity holds:
\begin{equation}
\begin{split}
    D_{a}^{4}\ln \Theta(\mathbf{z})+6(D_{a}^{2}\ln \Theta(\mathbf{z}))^{2}+ 
3D_{a}'D_{a}'\ln \Theta(\mathbf{z})
-2D_{a}D_{a}''\ln \Theta(\mathbf{z})&\\
-24 C_{1}D_{a}^{2}\ln 
\Theta(\mathbf{z})+12(10C_{2}-3C_{1}^{2})&=0
    \label{eq:fay}.
\end{split}
\end{equation}
Here the constants $C_{1}$, $C_{2}$ turn up in the Taylor expansion of 
the differential $d\Omega_{1}$ from above ($a=\infty^{+}$), 
\begin{equation}
    d\Omega_{1}=-\frac{1}{t^{2}}+2C_{1}-(6C_{1}^{2}-12C_{2})t^{2}+\ldots
    \label{eq:c12}.
\end{equation}
We put $u = 2(D_{a}^{2}\ln \Theta(\mathbf{z}) -2C_{1})$ and get for 
relation (\ref{eq:fay}) after differentiation with $D_{a}^{2}$
\begin{equation}
    D_{a}(D_{a}^{3}u+6uD_{a}u-2D_{a}''u)+3D_{a}'D_{a}'u=0
    \label{eq:fay2}.
\end{equation}
Because of (\ref{eq:bperiods}) this equation is for 
$\mathbf{z}=\mathbf{U}x+\mathbf{V}y+\mathbf{W}t+\mathbf{D}$ 
equivalent to the KP equation. 


\subsection{Reduction to KdV equation and solitonic limit}

If the hyperelliptic surface $\Sigma_{g}$ is branched at infinity,
i.e.\ if $F_{g+1}\to\infty$ in (\ref{hyper1}), the integral
$\Omega_{2}$ is single valued. Thus, all its $a$- and $b$-periods
vanish which implies that there is no $y$-dependence in the formula
(\ref{eq:solution}). Then the KP equation reduces to the KdV
equation,
\begin{equation}
    6uu_{x}+u_{xxx}-4u_{t}=0
    \label{eq:kdv}.
\end{equation}
The hyperelliptic surface is now defined by the algebraic curve 
\begin{equation}
    \mu^{2}= (K-E_{g+1})\prod_{i=1}^{g}(K-E_{i})(K-F_{i})=:
    K^{2g+1}-S_{1}K^{2g}+S_{2}K^{2g}-\ldots +S_{2g+1}
    \label{eq:hyper2}.
\end{equation}
All definitions apply as before. We use a cut-system of the form 
Fig.~\ref{fig:cut-system} where all $b$-cycles start at the cut 
$[E_{g+1},\infty]$ along the positive real axis. 

As a local parameter at infinity we can use $\i\sqrt{\lambda}$. The
differentials of the second kind we need here read
\begin{equation}
    d\Omega_{1}= 
    \frac{\i d\lambda}{2\mu(\lambda)}(\lambda^{g}+c_{1}\lambda^{g-1}+\ldots 
    c_{g}), 
    \label{eq:kdvdiff}
\end{equation}
and 
\begin{equation}
    d\Omega_{3}=-\frac{3\i
d\lambda}{2\mu(\lambda)}(\lambda^{g+1}-\frac12 S_{1}\lambda^{g}+d_{1}
\lambda^{g-1}+\ldots+d_{g})
    \label{eq:kdcdiff2},
\end{equation}
where the constants $c_{i}$ and $d_{i}$ are again defined by the 
condition that all $a$-periods of the above differentials vanish. 
Writing the normalized holomorphic differentials as 
\begin{equation}
    d\omega_{n}=\sum_{k=1}^{g}\frac{c_{nk}\lambda^{g-k}}{\mu}d\lambda
    \label{eq:holo2},
\end{equation}
we get for the $b$-periods of the differentials of the second kind 
\begin{equation}
    U_{n}=c_{n1}, \quad W_{n}=c_{n2}+\frac{S_{1}}{2}c_{n1}
    \label{eq:bper2}.
\end{equation}
The constant $c$ reads 
\begin{equation}
    c = -\left(\frac{S_{1}}{2} + c_{1}\right)
    \label{eq:ckdv}.
\end{equation}

An interesting limiting case of the theta-functional solutions on a 
genus $g$ surface is the so-called solitonic limit, see 
\cite{algebro,mumford}. In this case 
$E_{i}\to F_{i}$ for $i=1,\ldots,g$ which leads to the $g$-soliton 
solution. Since $g$ of the cuts collapse to double points, the 
diagonal elements of the Riemann matrix diverge in the used cut-system
as $B_{ii}\sim 2\ln 
(F_{i}-E_{i})$ whereas all other $a$- and $b$-periods remain finite. 
The theta-series thus breaks down to a sum containing only elementary 
functions. To obtain the standard form of the $g$-soliton, we choose 
the vector $\mathbf{D}$ in (\ref{eq:solution}) to correspond to the 
half-integer characteristic 
\begin{equation}
    \frac{1}{2}\left[
    \begin{array}{ccc}
        1 & \ldots & 1  \\
        0 & \ldots & 0
    \end{array}
    \right]
    \label{eq:chars}.
\end{equation}

On an elliptic surface we get with the above relations for (\ref{eq:solution}) 
the 1-soliton solution of KdV equation, 
\begin{equation}
    u = \frac{U^{2}}{2\cosh^{2}\frac{z}{2}}+2c, \quad z = Ux+Wt,
    \label{eq:sol2}
\end{equation}
where  
\begin{equation}
    U = 2\sqrt{E_{2}-E_{1}}, \quad W = 
    -2\sqrt{E_{2}-E_{1}}S_{1},\quad 
    c =-\frac{E_{2}}{2}
    \label{eq:soliton1}.
\end{equation}
Similarly we get for the 2-soliton 
\begin{equation}
    u-2c = 
    \frac{U_{1}^{2}-U_{2}^{2}}{2}\frac{U_{1}^{2}\cosh^{2}
    \frac{z_{2}}{2}+U_{2}^{2}\sinh^{2}\frac{z_{1}}{2}}{(U_{1}\cosh\frac{z_{1}}{2}
    \cosh\frac{z_{2}}{2}-U_{2}\sinh\frac{z_{1}}{2}\sinh\frac{z_{2}}{2})^{2}}
    \label{eq:sol23},
\end{equation}
where 
\begin{equation}
    z_{i}= U_{i}x+W_{i}t, \quad U_{1}=2\sqrt{E_{3}-E_{1}}, \quad
    U_{2}=2\sqrt{E_{3}-E_{2}}
    \label{eq:pers1},
\end{equation}
and
\begin{equation}
    W_{1}=-\sqrt{E_{3}-E_{1}}(E_{3}+2E_{1}),\quad 
    W_{2}=-\sqrt{E_{3}-E_{2}}(E_{3}+2E_{2}), \quad c = -\frac{E_{3}}{2}.
    \label{eq:pers2}
\end{equation}
A generalization to higher genus is straight forward, see 
\cite{soliton}. Formulas (\ref{eq:soliton1}) and (\ref{eq:sol23}) 
hold also for the KP equation with the appropriate definition of 
$\mathbf{z}$ and $\mathbf{U}$.

\section{Numerical implementations}
\label{sec:spectral}

The numerical task in this work is to approximate and evaluate
analytically defined functions as accurately and efficiently as
possible. To this end it is advantageous to use (pseudo-)spectral
methods which are distinguished by their excellent approximation
properties when applied to smooth functions. Here the functions are
known to be analytic except for isolated points. In
this section we explain the basic ideas behind the use of spectral
methods and describe in detail how the theta-functions and its 
derivatives can be obtained to a high degree of accuracy.

\subsection{Spectral approximation}

The basic idea of spectral methods is to approximate a given function
$f$ globally on its domain of definition by a linear combination
\[
f \approx \sum_{k=0}^N a_k \phi_k,
\]
where the functions $\phi_k$ are taken from some class of functions
which is chosen appropriately for the problem at hand. 
The spectral coefficients are determined by a 
so-called collocation method, i.e.\ by the condition
that $f(x_l) = \sum_{k=0}^N a_k
\phi_k(x_l)$ at selected points $(x_l)_{l=0:N}$. Since we are 
interested in an approximation of functions defined on a finite 
interval, orthogonal polynomials, in particular Chebyshev  $T_n(x)$
polynomials will be chosen. They are defined on the interval 
$I=[-1,1]$ by the relation 
\[
T_n(\cos(t)) = \cos(n t), \text{where } x = \cos(t),\qquad t\in[0,\pi].
\] 
The Chebyshev polynomials
are used because they have very good approximation properties and
because one can use fast transform methods when computing the
expansion coefficients from the function values provided one chooses
the collocation points $x_l=\cos(\pi l/N)$ (see~\cite{fornberg} and
references therein).

Let us briefly summarize here some basic properties of Chebyshev
polynomials: 
The addition theorems for sine and cosine imply the recursion
relations
\begin{equation}
  \label{eq:recursderiv}
  \frac{T'_{n+1}(x)}{n+1} - \frac{T'_{n-1}(x)}{n-1} = 2 T_n(x)
\end{equation}
for their derivatives. The Chebyshev polynomials are orthogonal on $I$
with respect to the hermitian inner product
\[
\left< f, g \right> = \int_{-1}^1 f(x) \bar g(x) \,\frac{d x}{\sqrt{1-x^2}}.
\]
We have
\begin{equation}
  \label{eq:ortho}
  \left< T_m , T_n \right> = c_m \frac\pi2\, \delta_{mn}
\end{equation}
where $c_0=2$ and $c_l=1$ otherwise.

The fact that $f$ is approximated globally by a finite sum of
polynomials allows us to express any operation applied to $f$
approximately in terms of the coefficients. Let us illustrate this in
the case of integration. So we assume that $f = p_N =\sum_{n=0}^N a_n
T_n$, and we want to find an approximation of the integral for $p_N$,
i.e., the function
\[
F(x) = \int_{-1}^x f(s)\, ds,
\]
so that $F'(x)=f(x)$. We make the ansatz $F(x) = \sum_{n=0}^N b_n\,
T_n(x)$ and obtain the equation
\[
F' = \sum_{n=0}^N b_n\,T'_n = \sum_{n=0}^N a_n T_n = f.
\]
Expressing $T_n$ in terms of the $T'_n$ using~\eqref{eq:recursderiv}
and comparing coefficients  determines all $b_l$ in terms of the
$a_n$ except for $b_0$. This free constant is determined by the
requirement that $F(-1)=0$. 
Thus, to find an approximation of the definite 
integral $\int_{-1}^{1}f(x)dx$ we
proceed as described above, first computing the coefficients $a_n$ of
$f$, computing the $b_n$ and then calculating the sum of the necessary
coefficients.

\subsection{Numerical treatment of the periods}

The quantities entering formula~(\ref{eq:solution}) are the periods of
certain differentials on the Riemann surface. The value of the
theta-function is then approximated by a finite sum.

The periods of a hyperelliptic Riemann surface can be expressed as
integrals between branch points. Since we need in our example the
periods of the holomorphic differentials and differentials of the
second kind with poles at $\infty^{\pm}$, we have to consider integrals
of the form
\begin{equation}
    \int_{P_{i}}^{P_{j}}\frac{K^{n}dK}{\mu(K)}, \quad n=0,\ldots,g+3
    \label{period1},
\end{equation}
where the $P_{j}$, $j=1,\ldots,2g+2$ denote the branch points of
$\Sigma_{g}$.

We parametrize the straight line between the two branch points $P_i$
and $P_j$ by 
\[
K = \frac12 (P_j + P_i)  + \frac{t}2 (P_j - P_i)
\]
to obtain the integral~(\ref{period1}) in the normal form
\begin{equation}
\label{eq:int_aperiod} \int_{-1}^1 \frac{\alpha_0 + \alpha_1 t 
+\ldots +
\alpha_{g+3} t^{g+3}}{\sqrt{1-t^2}} \;H(t) \,dt,
\end{equation}
where the $\alpha_i$ are complex constants and where $H(t)$ is a
continuous (in fact, analytic) complex valued function on the interval
$[-1,1]$.  This form of the integral suggests to express the powers
$t^n$ in the numerator in terms of Chebyshev polynomials and to
approximate the function $H(t)$ by a linear combination of Chebyshev
polynomials
\[ 
H(t) = \sum_{n\ge0} h_n T_n(t).  
\] 
The integral is then calculated with the help of the orthogonality
relation~(\ref{eq:ortho}) of the Chebyshev polynomials. 

This is the procedure used in I for general position of the branch
points which can also be applied here for low genus. The problem is
that there does not seem to be a direct way to express the terms
$K^{n}$ in (\ref{period1}) after the linear transformation in terms of
Chebyshev polynomials in $t$. This is an obstacle to the determination
of the periods for general genus. Another problem are situations close
to the solitonic limit. Since the cut-system Fig.~\ref{fig:cut-system}
is adapted to this case, the $a$-periods do not lead to problems. For
the $b$-periods, the function $H$ in the expression for
$b_{i}$~(\ref{eq:int_aperiod}) behaves like $1/\sqrt{t+E_{i}-F_{i}}$
near $t=0$. For $E_{i}\sim F_{i}$ this behavior is only satisfactorily
approximated by a large number of polynomials. The use of a large
number of polynomials however does not only require more computational
resources but has the additional disadvantage of introducing larger
rounding errors. It is therefore worthwhile to reformulate the problem
since a high number of polynomials would be necessary to obtain
optimal accuracy in this case.

It is possible to address both problems in one approach. The idea is
to use substitutions in the integrals (\ref{period1}) leading to a
regular integrand. To determine the $a$-periods, we use
\begin{equation}
    K = \frac{E_{i}+F_{i}}{2}+\frac{F_{i}-E_{i}}{2}\cosh x
    \label{eq:apers}.
\end{equation}
After a linear transformation which transforms the integration path to
the interval $[-1,1]$, the integral is computed with the Chebyshev
integration routine sketched above. This also works in situations
close to the solitonic limit. To treat the $b$-periods in this case,
we split the integral from $F_{i}$ to $E_{i+1}$ in two integrals from
$F_{i}$ to $(F_{i}+E_{i+1})/2$ and from $(F_{i}+E_{i+1})/2$ to
$E_{i+1}$. In the former case we use the substitution (\ref{eq:apers})
and
\begin{equation}
    K = \frac{E_{i+1}+F_{i+1}}{2}+\frac{F_{i+1}-E_{i+1}}{2}\cosh x
    \label{eq:berps}
\end{equation}
in the latter. These substitutions lead to a regular integrand even in
situations close to the solitonic limit. After a linear
transformation, the integrals are computed with the Chebyshev
integration routine. The disadvantage of this approach is that the
definition of the square root adapted to the cut-system in I via
monomials cannot be used in the case of the above non-linear
transformations. Here the root is chosen in a way that the integrand
is a continuous function on the path of integration. This can be
achieved in general for real branch points.

\begin{figure}[htb]
    \centering \epsfig{file=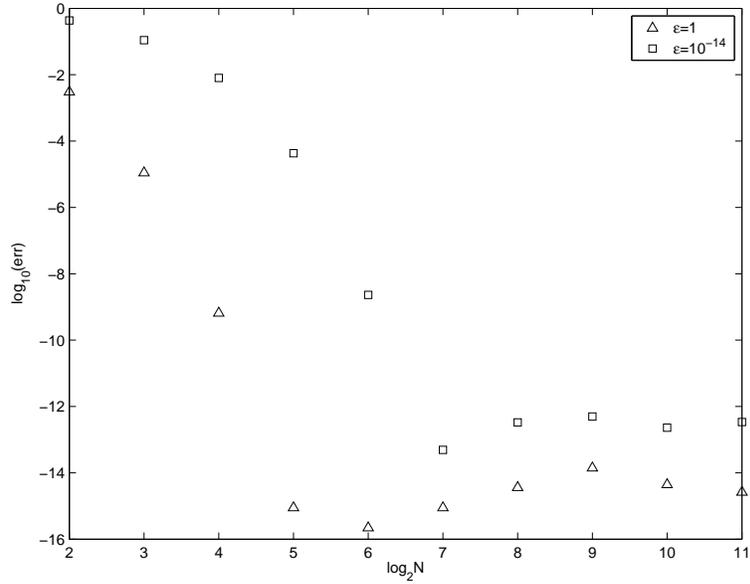,width=10cm}
    \caption{Test of the numerics for the $a$-periods at several 
    points in the space-time. The error is shown in dependence of the 
    number $N$ of Chebyshev polynomials.}
    \label{fig:kperr}
\end{figure}

To test the numerics we use the fact that in our case the integral of
any holomorphic differential along a contour surrounding the cut
$[E_{1},F_{1}]$ in positive direction is equal to minus the sum of all
$a$-periods of this integral. Since this condition is not implemented
in the code it provides a strong test for the numerics. The symmetry
of the Riemann matrix (the matrix of $b$-periods of the holomorphic
differentials after normalization) can be used to estimate the error
in the numerical evaluation of the $b$-periods. We define the function
$err$ as the maximum of the norm of the difference in the $a$-periods
discussed above and the difference of the off-diagonal elements of the
Riemann matrix. Additional tests for the periods $\mathbf{U}$,
$\mathbf{V}$ and $\mathbf{W}$ follow from (\ref{eq:bperiods}).  We
show in Fig.~\ref{fig:kperr} a genus 2 situation with the branch
points $[-3,-2,-1,-1+\epsilon,1,1+\epsilon]$.  It can be seen that 32
polynomials are sufficient in general position ($\epsilon=1$) to
achieve optimal accuracy. Since MATLAB works with 16 digits, machine
precision is in general limited to 14 digits due to rounding
errors. These rounding errors are also the reason why the accuracy
drops slightly when a higher number of polynomials is used.
In an almost degenerate situation ($\epsilon=10^{-14}$), optimal accuracy 
is achieved with 128 polynomials. Note that all quantities entering 
(\ref{eq:solution}) are within $10^{-14}$ and better equal to the limits 
given in (\ref{eq:pers1}) and (\ref{eq:pers2}). Thus the solitonic 
limit can be reached numerically with machine precision.

The above procedure to determine the periods can be directly
implemented within MATLAB for arbitrary genus. Due to the efficient
vectorization algorithms of MATLAB, the calculation of the periods
with optimal accuracy takes only a second even for genus 10 on the
used low-end computers. The limiting factor is here whether the matrix
$A$ of $a$-periods is ill conditioned. This is the case if a
considerable number of the entries of $A$ are of the order of the
rounding error ($10^{-16}$). Thus due to the limited number of digits,
the inversion of the matrix $A$ which is necessary to determine the
Riemann matrix can only be carried out with reduced accuracy which is
independent of the number of polynomials used in the spectral
expansion.  For large genus, this becomes the limiting factor in the
determination of the Riemann matrix. For instance in the case of the
genus 20 surface with branch points $[-21,20,\ldots,20]$, the identity
for $a$-periods is satisfied with 32 polynomials up to $10^{-15}$, the
symmetry of the Riemann matrix only up to $10^{-6}$ since the matrix
$A$ is badly scaled. The calculation of the periods takes less than a
second in this case.

\subsection{Theta-functions}

The calculation of the theta-functions below is applicable for 
arbitrary  Riemann surfaces, it is not limited to the hyperelliptic 
case. The theta-series~(\ref{theta}) for the Riemann theta-function (the 
theta-function in (\ref{theta}) with zero characteristic, 
theta-functions with characteristic follow from (\ref{theta2})) is
approximated as the sum
\begin{equation}
    \Theta(\mathbf{x}|\mathbf{B})
    =\sum_{n_{1}=-N}^{N}\ldots\sum_{n_{g}=-N}^{N}\exp\left\{
    \frac{1}{2}\langle \mathbf{M},\mathbf{ B}\mathbf{M}\rangle 
    +\langle \mathbf{M},\mathbf{x}\rangle\right\},
    \label{eq:thetasum}
\end{equation}
where $\mathbf{M}$ is a vector in $\mathbb{Z}^g$ with the components
$n_{1},\ldots, n_{g}$. We use the periodicity properties of the
theta-function (\ref{eq:periodicity}) to minimize the absolute value
of $x$. The value of $N$ is determined by the condition that terms in
the series~(\ref{theta}) for $n>N$ are strictly smaller than some
threshold value $\epsilon$ which is taken to be of the order of
$10^{-16}$. To this end we determine the eigenvalues of $\mathbf{B}$
and demand that
\begin{equation}
    N> -\frac{1}{gB_{max}}\left(||\mathbf{x}||+\sqrt{||\mathbf{x}||^{2}
    +2\ln \epsilon gB_{max}}\right)
    \label{eq:N},
\end{equation}
where $B_{max}$ is the real part of the eigenvalue with maximal real
part ($\mathbf{B}$ is negative definite). Similar formulas can be
obtained for theta-derivatives. For a more sophisticated analysis of
theta summations see~\cite{dubrovin97,deconinck03}. In the studied examples we
found values of $N$ between 2 and 40. If the eigenvalues of
$\mathbf{B}$ differ by more than an order of magnitude which can
happen close to partial degeneration of the surface, a summation over
an ellipse rather than over a sphere in the hypercube
$(n_{1},\ldots,n_{2g})$, i.e.\ different limiting values for the
$n_{i}$ as in~\cite{deconinck03} could be more efficient. The
summation over the hypercube has, however, the advantage that it can
be implemented in MATLAB for arbitrary genus. In addition it makes
full use of MATLAB's vectorization algorithms outlined below. Thus it
is questionable whether a summation over an ellipse would be more
efficient in terms of computation time in this setting.

Here we made use of MATLAB's efficient way to handle matrices. We
generate a $2N+1$-dimensional array containing all possible index
combinations and thus all components in the sum (\ref{eq:thetasum})
which is then summed. To illustrate this we consider the simple
example of genus 2 with $N=2$. The summation indices are written as
$(2N+1)\times (2N+1)$-matrices since $g=2$. Each of these matrices
contains $2N+1$ copies of the vector with integers
$-(2N+1),\ldots,2N+1$. $N_{2}$ is the transposed matrix of
$N_{1}$. Explicitly, we have
\begin{equation}
    N_{1} = \left(
    \begin{array}{ccccc}
        2 & 2 & 2 & 2 & 2  \\
        1 & 1 & 1 & 1 & 1  \\
        0 & 0 & 0 & 0 & 0  \\
        -1 & -1 & -1 & -1 & -1  \\
        -2 & -2 & -2 & -2 & -2
    \end{array}
    \right),\quad 
    N_{2} = \left(
       \begin{array}{ccccc}
           2 & 1 & 0 & -1 & -2  \\
           2 & 1 & 0 & -1 & -2  \\
           2 & 1 & 0 & -1 & -2  \\
           2 & 1 & 0 & -1 & -2  \\
           2 & 1 & 0 & -1 & -2
       \end{array}
       \right)
    \label{eq:indices}.
\end{equation}
The terms in the sum (\ref{eq:thetasum}) can thus be written in 
matrix form 
\begin{equation}
    \exp\left(\frac{1}{2}N_{1}\star N_{1}B_{11}+N_{1}\star N_{2}B_{12}+\frac{1}{2}
    N_{2}\star N_{2}B_{22}+N_{1}\star x_{1}+N_{2}\star x_{2}\right)
    \label{eq:thetasum2},
\end{equation}
where the operation $N_{1}\star N_{2}$ denotes that each element of
$N_{1}$ is multiplied with the corresponding element of $N_{2}$. Thus,
the argument of $\exp$ is a $(2N+1)\times (2N+1)$-dimensional
matrix. Furthermore, the exponential function is understood to act not
on the matrix but on each of its elements individually, producing a
matrix of the same size. The approximate value of the theta-function
is then obtained by summing up all the elements in
(\ref{eq:thetasum2}).

The most time consuming operations are the determination of the 
bilinear terms involving  the Riemann matrix. If one wants to calculate 
solutions to the KP equation, these terms only have to be determined 
once for a given surface. The integer $N$ is in this case fixed 
for the largest $||x||$ in the plot. We note that the summation is 
very fast even though the used determination of $N$ is rather crude. For 
instance in the case of genus $2$  with $N=100$, the calculation of 
the theta-function takes 0.1s on the used computers.

The limiting factor is here the available memory since arrays of the
order $(2N+1)^{g}$ have to be multiplied with each other. On the used
low-end computers we could deal with rather general genus 4
situations, but the limit was reached for genus 6 and $N=5$. The
summation is still very efficient, the calculation of the bilinear
terms and the determination of the coefficients took 16s in the latter
case, the subsequent calculation of the linear terms in
(\ref{eq:thetasum2}) and the summation, which have to be carried out
for each value of $x$ and $t$, took roughly 4s. Thus the limitations
we had to face were not due to the computing time but due to missing
memory.


The quality of the theta summation can be checked directly by putting
the approximate solution into the differential equation. The necessary
derivatives of the theta-function entering the function $u$ and its
derivatives can be calculated directly, e.g.\
\begin{equation}
    \partial_{x}^{n}\Theta(\mathbf{z},\mathbf{B}) = 
    \sum_{n_{1}=-\tilde{N}}^{\tilde{N}}\ldots\sum_{n_{g}=-\tilde{N}}^{\tilde{N}}
    (\langle \mathbf{M},\mathbf{U}\rangle)^{n}\exp\left\{
        \frac{1}{2}\langle \mathbf{M},\mathbf{ B}\mathbf{M}\rangle 
        +\langle \mathbf{M},\mathbf{z}\rangle\right\}
    \label{eq:thetader},
\end{equation}
where $\tilde{N}$ is determined as $N$ for the theta-series.
The differential equations could be satisfied to the order of $10^{-13}$ 
and better. For the plots in the next section we use an accuracy of 
the order of $10^{-10}$. The error is mainly due to the fact that we 
did not choose $N$ big enough that also the terms in 6th derivative of the 
form (\ref{eq:thetader}) are of the order of the rounding error. 
For the solution only the second derivative is important. The accuracy 
of the solution is thus in general much higher as can be seen for 
instance in the solitonic example where the differential 
equation is satisfied to the order of $10^{-13}$, but the difference 
to the solitonic solution is smaller than $10^{-15}$.

\section{Hyperelliptic solutions to KdV and KP equations}
\label{sec:examples}

In this section we present plots of hyperelliptic solutions to the KdV
and the KP equation as examples for the previously discussed
code. Such plots can already be found in \cite{mumford,bobbor} and
\cite{dubrovin97}. We will show general situations as well as almost
degenerate surfaces which are identical to their corresponding
solitonic solution up to numerical accuracy. In all plots, the vector
$\mathbf{D}$ in (\ref{eq:solution}) is chosen to correspond to the
characteristic (\ref{eq:chars}).

It turns out that almost solitonic solutions take less computational 
resources. The calculation of the periods requires the use of more 
polynomials, but the most time-consuming part is the theta summation. 
Since the diagonal elements of the Riemann matrix diverge in this 
limit, less terms have to be considered in the summation. The 
generation of the plots is thus considerably faster than in general 
position of the branch points.

\begin{figure}[htb]
    \centering \epsfig{file=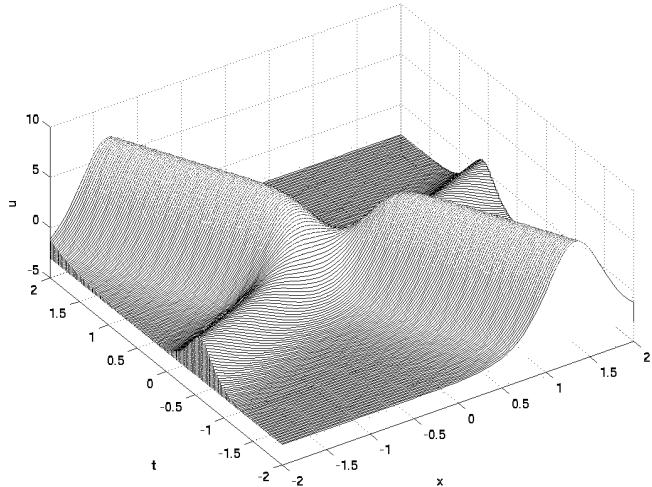,width=10cm}
    \caption{`Almost' solitonic genus 2 solution to the KdV equation.}
    \label{fig:kdv2s}
\end{figure}

To begin we want to discuss plots of genus 2 solutions to the KdV
equation. We consider a hyperelliptic surface of the form
(\ref{eq:hyper2}) with branch points
$[-2,-2+\epsilon,-1,-1+\epsilon,3]$. In Fig.~\ref{fig:kdv2s} we show
the case for $\epsilon=10^{-14}$ which is identical to the $2$-soliton
solution within machine precision. The difference between the shown
plot and the solution (\ref{eq:sol23}) is actually less than
$10^{-15}$. The one dimensional waves in shallow water are depicted in
dependence on $x$ and $t$. It can be seen that a soliton coming from
the right and traveling in negative $x$-direction has a collision with
a soliton traveling in positive $x$-direction. At the collision the
typical phase shift can be observed, otherwise the solitons are
unaffected.

\begin{figure}[htb]
    \centering \epsfig{file=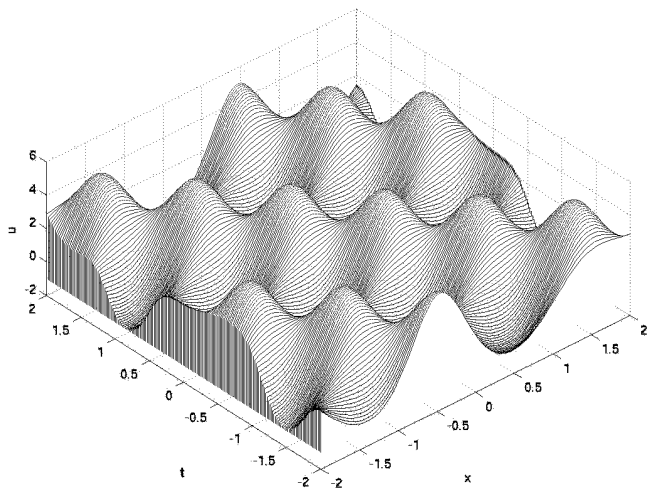,width=10cm}
    \caption{Almost periodic genus 2 solution to the KdV equation.}
    \label{fig:kdv2}
\end{figure}

In Fig.~\ref{fig:kdv2} we show the case $\epsilon=1$. The almost 
periodic nature of the solution is clearly recognizable. The solution 
can be interpreted as an infinite train of solitons. 

\begin{figure}[htb]
    \centering \epsfig{file=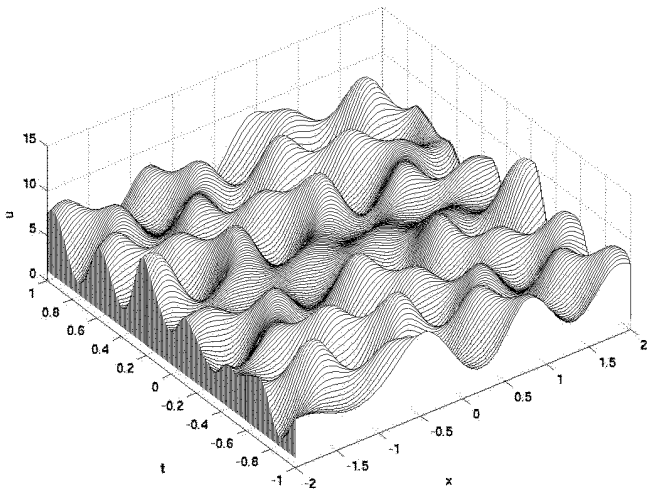,width=10cm}
    \caption{Almost periodic genus 6 solution to the KdV equation.}
    \label{fig:kdv4}
\end{figure}

\begin{figure}[htb]
    \centering \epsfig{file=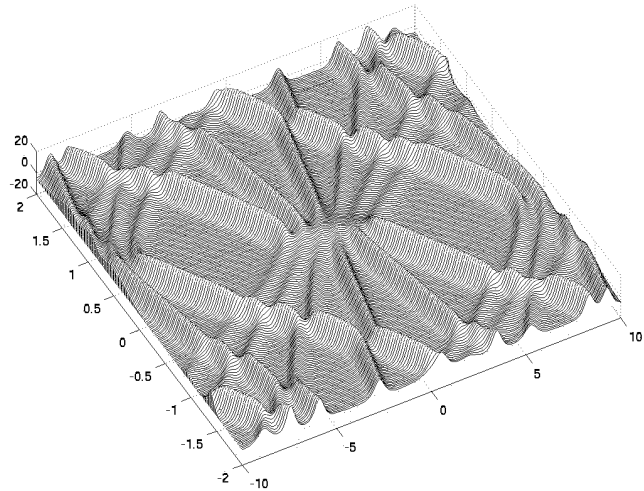,width=10cm}
    \caption{Almost solitonic genus 6 solution to the KdV equation.}
    \label{fig:kdv6s}
\end{figure}

To obtain solutions on a surface of genus 6  we considered the 
surface with the branch points $[-6,-6+\epsilon,-4,-4+\epsilon,
-2,-2+\epsilon,0,ep,2,2+\epsilon,4,4+\epsilon,6]$. 
The situation for $\epsilon=1$  is shown 
in Fig.~\ref{fig:kdv4}. The depicted situation is at the limit imposed 
by memory on the used computer.
In Fig.~\ref{fig:kdv6s} we show a solution of genus 6 in an almost 
solitonic situation. One can see the collision of 6 solitons at the 
center of the plot. 

\begin{figure}[htb]
    \centering \epsfig{file=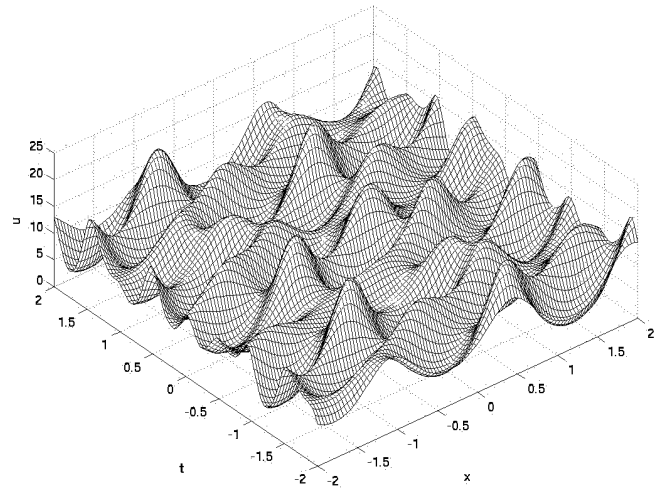,width=10cm}
    \caption{Almost periodic genus 4 solution to the KP equation.}
    \label{fig:kp4}
\end{figure}

\begin{figure}[htb]
    \centering \epsfig{file=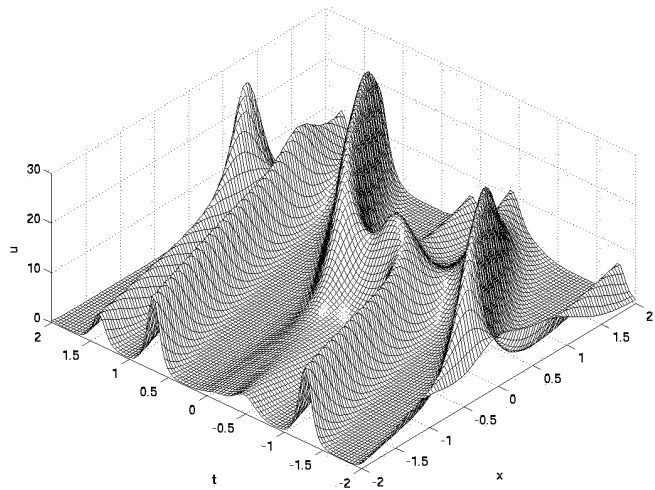,width=10cm}
    \caption{Almost solitonic genus 4 solution to the KP equation.}
    \label{fig:kp4s}
\end{figure}

A genus 4 solution of the KP equation is shown for fixed $t$ on the hyperelliptic 
surface with branch points $[-5,-4,-3,-3+\epsilon,
-1,-1+\epsilon,1,1+\epsilon,3,3+\epsilon]$. In Fig.~\ref{fig:kp4} we 
show the almost periodic situation for $\epsilon=1$. 
The almost solitonic case with $\epsilon=0.001$ is shown in 
Fig.~\ref{fig:kp4s}.

\section{Conclusion}
\label{sec:concl}
In this article we have presented a scheme based on spectral methods
to treat hyperelliptic theta-functions of in principle arbitrary genus
numerically. It was shown that an accuracy of the order of machine
precision could be obtained with an efficient code. As shown, spectral
methods are very convenient if analytic functions are
approximated. Close to singularities such as the degeneration of the
Riemann surface, analytic techniques must be used to obtain analytic
integrands as in the discussed example. This made it possible to study
the solitonic limit numerically within machine precision.

To consider the case that more than 2 branch points coincide, a higher
number of polynomials or additional analytical techniques have to be
used. As shown in I this makes it possible to study the parameter
space of hyperelliptic surfaces including almost degenerate cases with
at most 512 polynomials with maximal accuracy (here we used at most
128 polynomials). Both the calculation of the periods and the theta
summation are very efficient. The limiting factor for the periods is
the finite number of digits (16 in MATLAB) which leads to a drop of
accuracy if the matrix of $a$-periods is ill-conditioned. The theta
summation is limited by the available memory.

The presented numerical code is able to treat general hyperelliptic 
surfaces. The used method can in principle be extended to more 
general Riemann surfaces if the differentials are known explicitly as 
e.g.\ in the case of $Z_{N}$-curves.

\section*{Acknowledgment}
We thank A.~Bobenko and  D.~Korotkin for helpful discussions and hints. CK is
grateful for financial support by  the Schloessmann foundation.

\end{document}